# THE NETWORKS ARE OUT THERE: BUILDING CULTURAL AND ECONOMIC RESILIENCE THROUGH INFORMAL COMMUNITIES OF PRACTICE

Eduardo Marisca

Massachusetts Institute of Technology
77 Massachusetts Avenue, Building E15-331
Cambridge, MA, 02139, USA
e-mail: emarisca@mit.edu

**ABSTRACT**

This paper explores the possibilities offered by informal communities of practice to operate as "prototyping spaces" for innovation in the context of developing economies. It begins by looking at the concept of "economic complexity" and how it is useful in both guiding the priorities and evaluating the challenges developing economies face when attempting to drive growth and build measures of resilience, and raises the question of how these economies can both introduce higher complexity activities while compensating for latecomer disadvantages versus more complex economies. It then examines in detail the case of the Twin Eagles Group, a Peruvian video game development community in the 1990s, and how they reverse engineered technologies and global practices to pursue their creative objectives. Based on this case, it concludes by laying out some of the methodological challenges associated to researching this sort of community because of the multi-sited nature of their activities.

**INTRODUCTION**

Transforming developing economies into future-proof operations has proven to be a consistent challenge for decades. Going from a productive matrix built chiefly on primary activities including agriculture and resource extraction, to a higher value-added matrix of productive activities, is a resource-intensive process taking very long periods of time. It is not uncommon that precarious institutional environments and changing political and economic landscapes can derail large-scale commitments to introduce and grow new economic sectors that generate higher investment returns and reduced negative externalities.

These large-scale commitments usually require multi-level interventions designed to bridge the gap separating developing economies from their industrial and post-industrial competitors, and therefore not to compensate for several years or decades of latecomer disadvantage to achieve significant "competitive advantage". These efforts can involve some combination of government intervention, private sector development, technology investment, and education reform – any of which is a slow moving, complex system. At the same time, these commitments are both urgent and important: latecomer disadvantages only grow larger over time, and the need for accelerated industrialization pushes developing economies into a "race to the bottom" to cheapen their labor and relax their policing of regulation to encourage foreign investment.

The present paper is an initial attempt at exploring alternatives to this form of engagement – both to the absolute need for large-scale commitments, and to the costly compromises of "race to the bottom" scenarios. It builds on the assumption that a significant amount of creative and productive activity is going unnoticed by most indicators – even in these developing economies – because it flies outside the radar of economic production. Informal communities of practice are operating as innovation networks with growing creative output, but for various reasons are failing to articulate into economically viable sectors, or are failing to register within the attention of policymakers, researchers and investors.

These networks are out there and have been out there for a while, driven by self-motivated individuals with shared interests and passions and articulated through various forms of communications technologies making it possible, viable and sustainable for them to operate. Even in the absence of institutional climates favoring their emergence, or of access to resources facilitating their growth, these networks find the motivation to reverse engineer various layers of practices, institutions and technologies in order to accomplish their goals. In doing so, they're introducing skills and knowledge into their local contexts in ways that would otherwise be incredibly costly, and they're connecting with transnational communities of practice in attempting to do problem-solving and creative exploration.

In what follows, I will first go over the concept of economic complexity as the frame within which to understand the role these networks are playing. The work of Ricardo Hausmann and Cesar Hidalgo in this regard makes a solid case as to why developing economies need to focus increased efforts in diversifying their productive outputs as both a driver for accelerated growth and a measure to build resilience into otherwise fragile economies. But their work also points to how skill acquisition is one of the most complicated aspects of such diversification because of the large costs associated to it. I will argue that uncovering the communities and networks informally addressing these problems requires zooming in to the level of micro practices and social interactions that remain invisible to presently available metrics. Following that, I will briefly go over an example of one such network, operating almost invisibly, that was able to navigate these global and local processes with both successes and failures, and acquire and disseminate new skills in their local community: the Twin Eagles Group, a video game hacking and developing group operating in Lima, Peru, between 1989 and 2002, whose history is illustrative of the possibilities and challenges posed by the emergence of these networks in developing economies. Thirdly, I will highlight some of the methodological challenges facing the study of these networks because of their multi-sited character, and point to some of the questions researchers interested in these might want to consider when approaching them. I will finalize by providing some conclusions on the potential for the study of these networks and some questions for future exploration.

**THE PERKS OF BEING COMPLEX**

To understand the potential value these networks can contribute to an economy, it is first important to understand the larger networks to which they are connected – the networks of economic activity and productive output covering cities, countries and regions. Depending on the granularity or the "zoom level" at which one may choose to examine them, these networks will exhibit different characteristics and possibilities, and different configurations. And these configurations can be evaluated internally based on their diversity: how many different outputs an economy has. In Hausmann and Hidalgo's analysis, this becomes a measure of an economy's complexity (Hausmann and Hidalgo 2011).

Their comparative analysis of worldwide measures of complexity is based on trade data from recorded country exports. Based on this data, they can reconstruct a "product space" composed of all the possible products exported by some country in the world, and measure how various countries have a stake on some portion of that product space. If a country exports a product, then it follows that it produces such product. And if it produces such product, then it follows that its economy contains within it the necessary skills, knowledge and processes to make that production happen. Products can then be interpreted from the point of view of such skills, rather than in terms of the raw inputs required for production (as the raw inputs themselves can be interpreted in terms of the skills required to extract or produce them).

As a consequence of their analysis of this data, they encountered that even though all countries produce some products, most countries export products that are also exported by a large number of other countries. Countries exporting products exported by few other countries have higher indicators of prosperity, which follows from two different factors. First, products with fewer exporters imply more complex processes of production requiring more skills, and therefore represent a higher value in the market (because their complexity makes them both scarcer and more expensive). Second, a larger diversity of products within a national economy (measured by a larger diversity in product exports) translates to a wider variety of productive skills. Even though cars and airplanes are different products, they share some subset of overlapping skills in their production. A country producing cars will be closer to producing airplanes than one that doesn't. This means that as time goes on, countries with more skills are able to incorporate new productive activities faster than countries with fewer skills; and as economic landscapes shift, countries with more skills have a better chance at repurposing segments of their productive capacity in the face of changing demand than countries with fewer skills.

Though I find this analysis very compelling, there are a couple shortcomings to it worth pointing out, which are directly related to the limitations of the data itself. On the one hand, relying on international trade data presents a compelling picture of how products are circulated, but it fails to account for the ways in which services are provided across country lines, as service provision does not go through customs processing. Not only do services represent a significant contribution to an economy, but they too can also be understood in terms of the skills required to provide them. On the other hand, the nature of the data places the analysis at the national economy level, where multiple levels of granularity are otherwise possible: similar network configurations can be found at the city level, and even within at the neighborhood level, or in the other direction at the regional level. It becomes, however, much more complicated to perform such an analysis at these levels, as there is no consistent measurement of inputs and outputs. However, these shortcomings do not diminish at all the value of this type of analysis or of the concept of economic complexity for operating as a guideline in understanding growth and innovation within economies at any scale (whether we have available data or not). Countries are networked between each other, we can extrapolate, very much in similar fashion as cities and regions are, or conceptually, as sectors and industries are.

When thinking about economic complexity from the point of view of developing economies, we can begin to appreciate how developing economies struggle with their relatively low complexity. Not only do they tend to have less diverse exports, and therefore a presence in a smaller share of the product space, but those exports are also produced by a number of other countries, increasing competition between all exporters of a given product. Developing economies face the challenge of becoming more complex – that is, of producing a wider range of things other people want, and ideally have them be complex things that not many others produce.

This in turn raises two questions. First, how do countries (or economies of any scale) expand their command of the product space, that is, how do they begin to produce new things? Since products can be understood in terms of the skills required to produce them, this becomes a questions regarding skill acquisition and how new skills are introduced into an economy. But as Hausmann and Hidalgo point out, skill acquisition for non-complex economies becomes a huge issue because of enormous costs associated to it. Matters of infrastructure, education, policy, and so on all come into play when determining whether new skills can be successfully introduced in a sustainable way, and the presence of prior complex skills becomes an indicator of the ease of introducing new complex skills.

Second, how can economies that are able to successfully introduce new skills catch up to economies that were able to do so much earlier? Developing economies not only need to expand the diversity of their productive activities, but they need to do so fast and effectively enough that they will compensate for latecomer disadvantages. In practice, this means there is a requirement to innovate and diversify at a faster pace than complex economies have done so and continue to do so, but without the previous experiences and baseline layers of infrastructure complex economies have accrued in the process.

In the next section, I want to explore one possible answer to the problem of having to generate affordable and sustainable environments of innovation that can potentially turn into viable economic sectors. Because of the way in which economic output is measured within economies and firms – that is, in terms of direct productivity and economic output – there ends up being a remainder of activities that are not counted towards those outputs but account for a significant contribution towards making those outputs possible. According to Brynjolfsson and Saunders (2010), in the case of firms, significant pockets of innovation go unnoticed simply because available metrics fail to account for activities that do not directly contribute to expected outputs. Firms are, therefore, missing out on a lot of potential value already being generated within their bounds.

I believe a similar case can be made when looking outside of firms, at local or national economies: there are pockets of innovation and creativity out there that are going unnoticed because their activities fall outside the scope of available metrics and data – particularly because, in many cases, their activities are not considered to be economically productive at all. But these informal communities are generating the spaces and networks capable of introducing new skills into an economy even absent the prospects of immediate economic rewards. They are effectively creating prototyping spaces at a low cost, which may or may not be able to later become productive sectors and industries. These networks have been out there for a while, and they pose a challenge to us as researchers in terms of how to uncover and map them.

## **A CASE OF REVERSE ENGINEERING**

To explore this possibility, I want to go back in time several years and take a detailed look at a historical case that I think illustrates many of the challenges and opportunities of this approach. In between the late 1980s and the early 2000s, a group called the Twin Eagles Group (TEG) was very active in the Peruvian video game hacking, modding and development scene, claiming for themselves such landmarks as having the first Peruvian-made video game to be distributed internationally in the European market (*GunBee F-99*, 1999), and having the first Peruvian-made video game to be independently released in the local market (*The King of Peru 2*, 2002). Their history is a complex array of successes and failures as a small, shifting group of people negotiated their self-understanding as a group first as liberators of information, cracking software (initially, Commodore 64 software) to make it freely available to people; then as appropriators of global commodities, hacking games to make them more culturally meaningful for local audiences; and finally, as creators, trying to make and release their own games while making a living out of it.

The history of TEG is captured in various levels of detail in the group's website – a frozen online archive apparently last updated in 2005. When fully printed out (as I had to do on account of the frequent blackouts in availability), the archive spans 236 pages containing articles, game descriptions, member lists, event records, photographs, amongst other various things. The online archive also contains assorted related files such as disk images and playable files for some of their game projects.

Their story unfolds across an extremely curious timeline – one that could be rightfully labelled as "a series of unfortunate events". The group was active between 1989 and 2003, when it ultimately dissolved. This situates its inception during one of the toughest periods in recent Peruvian history: the later years of the first Garcia presidency were a social, political and economic disaster. In this environment, TEG was born as a Commodore 64 coding group, focused on making software available to the C64 community existing at the time in the city of Lima. While many other groups active at the same time were importing and selling C64 software for profit, sometimes going as far as stealing it from other groups or refusing to acknowledge due credit to crackers, TEG made no profit on the software they made available (although they would often charge depending on the amount of effort invested) and encouraged its distribution through then-nascent telephone Bulletin Board Systems (BBS) and self-organised "copy parties" where people could bring their own 5¼" disks and get their own pirated copies of software.

The neoliberal reforms of the Fujimori regime would affect them in various ways. TEG members were very active in the early BBS community, hosting two of the most interesting boards in the early nineties. Users would dial into a BBS using their regular phone line and connect directly to another computer for some time, during which they could download information or leave messages in the board while the call lasted, all of which was relatively cheap because of the low rates charged by CPT, the State-owned telephone company. When CPT was privatized in the early 90s and sold to the Spanish telecommunications giant Telefónica, calling rates rose dramatically and essentially crippled the BBS community. Simultaneously, political pressure to strengthen property rights led to an overhaul of the nation's intellectual property legislation in 1996, which was updated to contemplate newer media (such as software) and to facilitate enforcement by law agencies.

The group was active through the entirety of the Fujimori regime and the democratic transition between the years 2000 and 2001, where they attempted to regain some traction after long periods of inactivity and instability by releasing a series of games fitting the political theme of those years. However, their attempt in 2002 to release the first locally developed and published game ended with them getting into a complicated legal dispute with their distributor over royalties generated by the game (which is, of course, extremely ironic considering the group's origins), which at the time was being sold in physical CD-ROM format at retail establishments. In 2003, after releasing a highly questionable pornographic knock-off of *Tetris* (under the name *Samba de Oruga*) with the hopes of fundraising the money they needed to support their legal dispute, the group ultimately folded under legal and financial pressures.

The influence TEG had in the overall game development scene in Lima is very hard to map. Their production was certainly notable: according to their records, they released three games commercially (including the first Peruvian independent release, and the first Peruvian release in the European market), seven games as freeware, two game development code libraries and twenty-six hacked versions of console games. A significant share of this work was just happening through a process of reverse engineering: given the lack of formal training or available documentation, and of widespread access to information through the Internet, the only way for TEG members to understand how these technologies worked was through what would today seem like arcane methods. They would observe a piece of software functioning, and recreate pieces of code until they behaved exactly the same; or they would intervene in the normal operation of a program to observe what it was doing at the memory level, and inject operations and instructions at that very low level. The relative simplicity of computing platforms at the time played doubly to their favour: given the limited capacity of available technologies, the ceiling for what could be accomplished by smaller groups of programmers was within their reach; while at the same time, user expectations were considerably lower than what one would find today in the market.

Their reverse engineering operations were not strictly limited to the technological. At a micro level, TEG was also negotiating their inclusion into global practices of software development and of gaming culture. The ideals of the Free Software Movement were being crafted through the 80s, and the open-source operating system Linux would be first released in 1991, but the news about these developments had not reached the group when they were circumventing copy protection on software and openly distributing *warez* online – there is no mention of any of these in their records and, especially, in the issues of their early-90s *discmag*, *Smiling Panda*. But they were engaging with communities at the international level: their Amiga commercial release, *GunBee F-99*, was distributed and reviewed in Europe, and they have several records of communications with similar cracking groups in Europe, Mexico and Argentina. A significant part of their documents are available in both Spanish and English, and they actively maintained a dictionary file to help their developers acquaint themselves with technical terms in English. Moreover, their production can also be said to exemplify this reverse engineering of global practices: cracking a game to make it more locally meaningful was a similar practice taking place all over the continent as local communities creatively appropriated global cultural products, if only in small, self-contained ways.

Yet the fact remains that not only did TEG disappear from the gaming scene in the early 2000s, but it also failed to leave behind any enduring or significant legacy. Even more so, they could sometimes become divisive presence and perhaps even a significant obstacle to the articulation of such a community. Their rhetoric and underlying *ethos* was often a divisive presence probably hindering the articulation of a broader community of interest, instead playing to an essential dichotomy between crackers and *lamers*, the latter being a category used loosely to describe people who couldn't code, people who stole other's code, people who uses helper toolkits and gamemakers, and then anyone that generally disagreed with them and their methods. Their collected archive of Peruvian video games explicitly states: "We take in count videogames that were developed by programming them (in assembler or C, mainly). Games made with authoring tools like Flash, game-makers and level map editors will be NOT INCLUDED in this list because those tools do not promote the Investigation, Programming and Optimization knowledgment" (sic). Given that Assembler and C are amongst the hardest ways to introduce people to programming, it is hardly surprising that a strong community would not come together around such demands, or that the group would find it increasingly hard to recruit members as it clung to fading technologies such as the C64 and Amiga as generic PCs and video game consoles, which they regarded as inferior machines, began to gain the larger market shares.

There are two main reasons why I wanted to bring up the TEG case for discussion. The first one is that it provides an illustrative example of how an informal community of practice, operating from an economic, technological and cultural periphery, works to reverse engineer tools, processes and institutions to accomplish their objectives. Even lacking the access to the right infrastructure and systems that would enable them to produce their own video games – a lack that would've made most economic-centered endeavors unviable to begin with – they were still capable of finding the means, motives and opportunities to sustain an ongoing, continuously improving creative process, and to introduce into their locality a series of skills and knowledge that would otherwise remain unavailable. In this way, TEG was able to navigate what we saw earlier as one of the main challenges to increasing economic complexity in the context of developing economies: namely, the costs and complications associated to introducing new opportunities for effective skill acquisition.

The second reason is that it also works as a great example for the methodological challenges stemming from researching this sort of community and the networks weaved around it. If we agree that there's value in looking at the social practices taking place in these communities as a way to uncover networks of innovation operating informally outside the bounds of acknowledged economic activity, then we must also very closely consider the instruments and methods we're deploying to discover, map and understand them. Looking at TEG as a historical case required paying attention to historical elements happening in multiple locations and contexts simultaneously, and the field of research cannot be said to have been at any one location at all times. In the next section, I want to go back to this methodological observation and consider specifically some of the challenges researchers can confront in trying to map informal networks of innovation.

## **UNCOVERING NETWORKS: THREE METHODOLOGICAL CHALLENGES**

The TEG case is an example of a research context where a traditional qualitative approach looking at social practices and connections within a community or a network, especially an ethnographic one, is challenged by the lack of a stable, clearly bounded and localized field site. There was not specific site for TEG, even if there may have been a base of operations at some point: events, connections, influences, technologies, they were all happening and moving through a series of pathways connecting them all together. Their connections to other localities around the world, and their participation in cultures and norms that were adopting a global shape (such as video game and software development) present specific challenges to the researcher attempting to draw an account of their activities.

These challenges fall within a tradition, in the last few decades, of revising and questioning what the "object" and the "field" of research are in the case of qualitative research, and of ethnographic methods and instruments. As global and local relationships are renegotiated, so is the role and position of researchers towards the communities they're working with. I want to pay a closer look at three specific such challenges: the upsetting of the relationship between the global and the local, the operational difficulties this poses in terms of accessing a field site, and the expanded sense of reciprocity this access demands on the researcher.

### *Mapping the Global and the Local*

The notion of a multi-sited field of research is closely related to increasingly problematic notions of what it is to be "local" in the face of the emerging "global". The need for a multi-sited field begins to emerge "from anthropology's participation in a number of interdisciplinary (in fact, ideologically antidisciplinary) arenas that have evolved since the 1980s, such as media studies, feminist studies, science and technology studies, various strands of cultural studies, and the theory, culture, and society group." (Marcus 1995, 97) These newly available sites demanded new strategies of interrogation and analysis.

The shift to multi-sited fields upsets several of the conventions and assumptions – or "pre-theoretical commitments" in Henrietta Moore's description (2004) – of traditional qualitative research. Consequently, as documented by Marcus, it also creates a series of anxieties spawning directly from the fact that scientific guarantees come into question. What came into question was whether these field sites should be the domain of other disciplines or methods more attuned to geographical and historical diversity and complexity.

This does not automatically imply that the local is diluted out of existence, precisely because the global is both anywhere and nowhere at the same time. All experiences of the global are always through local instantiations. But the sense of what's local is now posed the challenge of accounting for its connections to what's happening in other places and landscapes.

Ethnography becomes a practice of tracing these accounts through multiple layers of meaning and practice (very much what it has done traditionally), but with the caveat that this tracing is not bounded to a place that is local in the geographical sense. Participant observation then needs to become one of the tools at the researcher's disposal in navigating these fields, but not the only one: researchers need to become acquainted with what Hugh Gusterson has termed "polymorphous engagement": that is, "interacting with informants across a number of dispersed sites, not just in local communities, and sometimes in virtual form; and it means collecting data eclectically from a disparate array of sources in many different ways." (Gusterson 1997, 39)

### *Getting Access*

The unbinding of the field site under the terms examined above poses an operational challenge. No longer can sites of social research be considered to be ready for the taking by the skilful researcher, because any group or community at any location finds itself now under multiple influences from external elements that are not uniquely situated at any given location. The same processes that have unsettled the former bounds of the traditional field site have altered the political status of researchers and their communities under study. Access to a multi-sited field is not only operationally complicated, requiring innovative methodological approaches, but is also culturally challenging and layered, and ceases to be about where the site is located or where the researcher is coming from. But even once access is gained, that doesn't necessarily imply one will get meaningful information to "follow" the practices of a community.

Yuri Takhteyev's ethnographic work with software developers in Rio de Janeiro (Takhteyev 2012) provides a good illustration. In his fieldwork in various software development projects, Takhteyev found that in order to understand the practices of developers working in Rio de Janeiro, it was necessary to look at how "worlds of practice" were constituted across locales. Not only technical knowledge, but also business practices, social norms and values and personal aspirations were circulated across these networks. This information only became available after taking an active stance as a collaborator in the software development projects he was observing. This shift yielded the realization of everything he was missing through traditional participant observation (or in any case, all those dimensions in which he was not participating): the codebase, the issue logs, the e-mail threads, the documentation, and so on. Even after gaining access to these locations, his position and commitment continued to be challenged by his informants (Takhteyev 2012, 16).

The people in these multi-sited fields are now in a position to have much clearer understanding of the role they play in global information flows (or "landscapes"), and consequently to impose much stricter demands on researchers trying to gain access to their networks – especially when those networks are directly connected to their livelihood.

*Reciprocating*

These instances of going beyond the traditional role of researcher could be understood as variations on the theme of participant observation: in a multi-sited field, the spectrum of activities one must participate in simply becomes more diverse. But again, there are epistemological and methodological issues at stake here. We can either hold on to positivistic expectations of ethnographic work, or we can alternatively acknowledge an expanded role of researchers not just as documenting but also contributing to define the field itself through their actions. While the decision of one over the other remains ultimately with the researcher, it is important for us to acknowledge the claims that underlie each position and the trade-offs we're incurring in when negotiating these boundaries.

Takhteyev's fieldwork experience, mentioned above, provides one illustration of such expanded sense of reciprocity. In dealing with members of the community he was researching, he was only able to gain further access to invisible locations by becoming an active contributor to the codebase. The burden in this case is significant, as his time on the field was now split between his commitments to his research project, and those to the development tasks he was volunteering to complete. At the same time, this possibility was only afforded to him because he already had knowledge of programming and software development practices, even if he was not fully fluent in them. We can legitimately ask whether this is the case for all researchers in all situations, or whether a researcher's interest and capacity to access a field now becomes mediated through professional competencies, subject-matter expertise or technical proficiency. Where traditional research sites were found to be valuable because of their otherness, multi-sited fields would appear in this sense to discriminate based on proximity rather than remoteness.

A different involvement with production processes can be found in Ian Condry's work with anime producers in Japan. In his case, late in his fieldwork, he became unexpectedly drafted to contribute to a part of the production process by doing some voice work for an anime movie he was following (Condry 2013, 144). It can hardly be said that had he not taken part of this production stage, the ultimate result would not have been accomplished. But these are instances in which researchers were forced by circumstances in the field to very explicitly take a stand around specific issues, practices or groups. The issue of positioning acquires greater importance when we're considering multi-sited fields because the researcher might be interacting not only with groups that are internally inconsistent, but also with multiple parties spread across a field whose actions might be directly in conflict or contradiction. The question over "whose side you're on" becomes more complicated the more sides are available to choose from.

Some contexts and political positions will lend themselves to more complicated evaluations of the allegiances of the researcher. If researchers are going to be expected to assume one position or another when dealing with the various locations in a field, anticipating this expectation and having solid arguments as to why one stands for or against something, and how far one is willing to go for such position, become useful rules of thumb. While occasional roles in production might seem relatively uncontroversial, larger roles in leadership or organisation might be more troubling – even more so in the case of advocacy or championing roles.

## CONCLUSIONS

In the early 1990s, Michael Porter presented a reading of innovation that sought to shift the explanations of technology change from the macroeconomic view to the microeconomic decisions and relations being established by firms in the market. In his view, available explanations were incapable of addressing "why and how meaningful and commercially valuable skills and technology are created" (Porter 1998, 9). This in turn led him to present an account of how firms benefitted from collaborating in clusters, as knowledge spillovers from individual firms were beneficial to the cluster as a whole, and became further incentives to skill development and technological improvement as a more or less organic process.

But such cluster building efforts involve large-scale, long-term commitments which themselves require baseline institutional frameworks and infrastructure availability which are many times unavailable in the context of developing economies. At the same time, failure to introduce such new skills and capabilities for technical change only increases the gap with more complex economies, and the difficulties of overcoming latecomer disadvantages when entering higher complexity markets. The work of Hausmann and Hidalgo also shows that commitment to a limited number of productive clusters, regardless of their complexity and added value, is not a good mechanism to provide resilience in the face of changing economic conditions, as smaller footprints within the product space imply fewer opportunities for repurposing skills and capabilities if an economy should find itself in the need to do so.

I've attempted to present an interpretation of informal communities of practice, as illustrated by the TEG case, as providing spaces for creative prototyping driven by self-motivation and social relations rather than by economic interest and expectation of financial payoffs. Networks of innovation operating in this fashion are flying under the radar of conventional metrics precisely because of this, yet within their activities they're coming up with creative ways to reverse engineer global processes and practices and adapt them creatively to the needs of local practitioners and audiences. These creative networks are out there and are failing to be considered when it comes to policy initiatives, long-term investment strategies, and even research projects.

At the same time, I've also attempted to express how these networks pose specific research challenges if we're interested in uncovering and understanding them. Because of their nature, their practices and potential require different approaches – for example, mapping their operations through various forms of qualitative research rather than through analysis of economic indicators – and these approaches themselves face methodological challenges from having to deal with multi-sited fields of research where activities are scattered through multiple localities, drawing together networks of both human and non-human actors (Latour 2005).

While these networks are actively introducing new skills into societies and economies, they often remain economically invisible and implausible. A qualitative approach focused on the social practices of these communities can help uncover their potential value and point towards opportunities for unblocking and developing that potential value. For many developing economies, unlocking the value in otherwise underestimated creative industries can provide hugely valuable pathways towards achieving much needed economic diversity and complexity.

Paying attention to these networks can provide fertile grounds on which to find forms of peripheral and grassroots innovation, as they figure out ways to circumvent the various obstacles and absences that would otherwise stop them from being effective. Their very existence highlights system deficiencies blocking full development of creative value. Some networks will thrive and some will wither, but out of their experiments new generations of makers and creators are emerging to break the ground on new forms of production in entirely unexpected places.

## **REFERENCES**

Brynjolfsson, Erik, and Adam Saunders. 2010. *Wired for Innovation : How Information Technology Is Reshaping the Economy*. Cambridge, Mass.: MIT Press.

Condry, Ian. 2013. *The Soul of Anime: Collaborative Creativity and Japan's Media Success Story*. Experimental Futures. Durham and London: Duke University Press.

Gusterson, Hugh. 1997. "Studying Up Revisited." *PoLAR: Political and Legal Anthropology Review* 20 (1): 114–119. doi:10.1525/pol.1997.20.1.114.

Hausmann, Ricardo, and Cesar A. Hidalgo. 2011. *The Atlas of Economic Complexity: Mapping Paths to Prosperity*. Cambridge, Mass: Center for International Development, Harvard University : Harvard Kennedy School : Macro Connections, MIT : Massachusetts Institute of Technology.

Latour, Bruno. 2005. *Reassembling the Social: An Introduction to Actor-network-theory*. Clarendon Lectures in Management Studies. Oxford ; New York: Oxford University Press.

Marcus, George E. 1995. "Ethnography In/of the World System: The Emergence of Multi-Sited Ethnography." *Annual Review of Anthropology* 24 (1): 95–117. doi:10.1146/annurev.an.24.100195.000523.

Moore, Henrietta L. 2004. "Global Anxieties: Concept-Metaphors and Pre-Theoretical Commitments in Anthropology." *Anthropological Theory* 4 (1) (March 1): 71–88. doi:10.1177/1463499604040848.

Porter, Michael E. 1998. *The Competitive Advantage of Nations: With a New Introduction*. Basingstoke: Macmillan.

Takhteyev, Yuri. 2012. *Coding Places: Software Practice in a South American City*. Acting with Technology. Cambridge, MA: MIT Press.